\documentclass[12pt]{article}
\usepackage[latin1]{inputenc}
\usepackage{graphicx}
\usepackage{color}
\usepackage{geometry}  
\usepackage{parskip}   
\usepackage{amsmath, amssymb} 
\usepackage{setspace} 
\usepackage{booktabs} 
\usepackage[round]{natbib} 

\newcommand{\bs}[1]{\ensuremath{\boldsymbol{#1}}} 

\begin{document}

\title{Assessing the Calibration of Subdistribution Hazard Models in Discrete Time} 
\author{Moritz Berger and Matthias Schmid}
\date{\today}

\maketitle

\begin{abstract}
\noindent
The generalization performance of a risk prediction model can be evaluated by its calibration, which measures the agreement between predicted and observed outcomes on external validation data. Here, methods for assessing the calibration of discrete time-to-event models in the presence of competing risks are proposed. The methods are designed for the class of discrete subdistribution hazard models, which directly relate the cumulative incidence function of one event of interest to a set of covariates. Simulation studies show that the methods are strong tools for calibration assessment even in scenarios with a high censoring rate and/or a large number of discrete time points. The proposed approaches are illustrated by an analysis of nosocomial pneumonia.
\end{abstract}


\section{Introduction} \label{sec:intro}

Over the past decade, risk prediction models have become an indispensable tool for decision making in applied research. Popular examples include models for diagnosis and prognosis in the health sciences, where risk prediction is used, e.g., for screening and therapy decisions \citep{steyerberg2009, moons2012I, liu2014} and models for risk assessment in ecological research, which have become an established tool to quantify and forecast the ecological impact of technology and development \citep{gibbs2011}.

A key aspect in the development of risk prediction models is the validation of generalization performance. This task, which is usually performed by applying a previously derived candidate prediction model to one ore more sets of independent external validation data, has been subject to extensive methodological research \citep{moons2012II, steyerberg2014, harrell2015, steyerberg2016, alba2017}. As a result, strategies for investigating the discriminative power (measuring how well a model separates cases from controls), calibration (measuring the agreement between predicted and observed outcomes) and prediction error (quantifying both discrimination and calibration aspects) of prediction models have been developed \citep{steyerberg2010}. Alternative techniques that additionally involve decision analytic measures include, among others, net benefit analysis \citep{vickers2016}, decision curve analysis \citep{vickers2006} and relative utility curve analysis \citep{baker2009, kerr2017}.

The aim of this article is to develop a set of methods for assessing the calibration of a prediction model with a time-to-event outcome. This class of models has been dealt with extensively during the past years, see, for example, \citet{henderson2005, witten2010, soave2018, braun2018}.
Here, we explicitly assume that event times are measured on a discrete time scale $t=1,2, \ldots$ \citep{ding2012, Tutz2016, Berger2018}, and that the event of interest may occur along with one or more ``competing'' events \citep{fahrmeir1996, Fine1999, lau2009, beyersmann2011, austin2016, lee2018}. Scenarios of this type are frequently encountered in observational studies with a limited number of fixed follow-up measurements, for instance in epidemiology \citep{andersen2012}. Since the study design does not allow for recording the exact (continuous) event times, it is only known whether or not an event of interest (or a competing event) occurred between two consecutive follow-up times $a_{t-1}$ and $a_t$, implying that the discrete time scale $t=1,2,\dots$ refers to a special case of interval censoring with fixed intervals. 

In recent years, several authors have developed measures and estimators for analyzing the generalization performance of discrete time-to-event models. For example, discrimination measures for discrete time-to-event models were proposed by \citet{Schmid2017}. Measures of prediction error were considered in \citet{Tutz2016}, Chapter 4. Graphical tools for assessing the calibration of discrete time-to-event predictions (not accounting for the occurrence of competing events) were explored in \citet{Berger2018}. Methods for assessing the generalization performance of discrete cause-specific hazard models (a common approach for competing risks analysis) have been recently proposed by \citet{heyard2019}.

We propose to base the calibration assessements for discrete competing risks models on the {\em cumulative incidence function} $F_j(t|\bs{x}) := P(T\le t , \epsilon = j | \bs{x})$, denoting by $T$ the time to the first event, by $\bs{x}$ a set of covariates, and by $\epsilon \in \{ 1, \ldots , J\}$ a random variable that indicates the occurrence of one out of $J$ competing events at $T$ \citep{Fine1999, klein2005}. In the following it will be assumed w.l.o.g.\@ that the event of interest and its cumulative incidence function are defined by $\epsilon = 1$ and $F_1(t|\bs{x})$, respectively. A popular method to derive predictions of $F_1(t|\bs{x})$ from a set of training data is to fit a proportional subdistribution hazard model \citep{Fine1999}. This approach, which is designed for the analysis of right-censored event times and which will also be considered here, has been recommended to analysts ``whenever the focus is on estimating incidence or predicting prognosis in the presence of competing risks'' \citep{austin2016}. While the original model proposed by  \citet{Fine1999} assumed the event times to be measured on a continuous time scale, the methods developed in this article are based on a recent extension of the subdistribution hazard modeling approach to discrete-time competing risks data \citep{berger2018sub}. Specifically, the model proposed by \citet{berger2018sub} is designed for estimating the {\em discrete subdistribution hazard} $\lambda_1 (t| \bs{x}) := P(T=t, \epsilon = 1 | (T \ge t) \cup (T \le t-1 , \epsilon \ne 1) , \bs{x})$, which defines a one-to-one relationship with the discrete cumulative incidence function $F_1 (t |\bs{x})$ (see Sections \ref{sec:notation} and \ref{sec:modelFitting} for details). The calibration of a subdistribution hazard prediction model may thus be characterized by how well the subdistribution hazards {\em observed} in a validation sample can be approximated by the respective {\em predicted} subdistribution hazards that are obtained from applying the prediction model to the validation data.

The proposed methodology for assessing the calibration of a discrete-time subdistribution hazard model comprises two parts, which both build on methods for binary regression: The first part (presented in Section~\ref{sec:calplot}) will be concerned with the derivation of an appropriate {\em calibration plot} that visualizes the agreement between the predicted and the observed subdistribution hazards. In the second part (Section~\ref{sec:logistic}), we will propose a {\em recalibration model} for discrete-time subdistribution hazard models that can be used to analyze calibration-in-the-large and refinement (i.e., the bias and the variation, respectively, of the predicted subdistribution hazards) along the lines of \citet{cox1958} and \citet{miller1993}. As will be shown in Sections \ref{sec:calplot} and \ref{sec:logistic}, the weights used in the subdistribution hazard modeling approach proposed in \citet{berger2018sub} allow for defining appropriate versions of the observed and predicted hazards (to be depicted in the calibration plot) and for fitting a weighted logistic recalibration model (giving rise to point estimates and hypothesis tests on calibration-in-the-large and refinement).

The proposed calibration assessments will be illustrated using a simulation study (Section \ref{sec:sim}) and the analysis of a data set on the duration to the development of nosocomial pneumonia (NP) in intensive care patients measured on a daily basis (Section \ref{sec:app}). Section \ref{sec:conc} summarizes the main findings of the article.

\section{Discrete subdistribution hazard models} \label{sec:notation}

Let $T_i$ be the event time and $C_i$ the censoring time of an i.i.d.\@ sample with $n$~individuals $i=1,\hdots,n$. Both $T_i$ and $C_i$ are assumed to be independent random variables taking discrete values in $\{1,2,\hdots,k\}$, where $k$ is a natural number. It is further assumed that the censoring mechanism is non-informative for $T_i$, in the sense that $C_i$ does not depend on any parameters used to model the event time \citep{kalbfleisch, kleinbaum}. For instance, in longitudinal studies with fixed follow-up visits, the discrete event times $1,\hdots,k$ may refer to time intervals $[0,a_1), [a_1, a_2), \hdots, [a_{k-1}, \infty )$, where $T_i=t$ means that the event has occurred in time interval $[a_{t-1},a_t)$ with $a_k=\infty$. For right-censored data, the time period during which an individual is under observation is denoted by $\tilde{T}_i = \min(T_i,C_i)$, i.e.\@ $\tilde{T}_i$ corresponds to the true event time if $T_i \leq C_i$ and to the censoring time otherwise. The random variable $\Delta_i:=I(T_i \leq C_i)$ indicates whether $\tilde{T}_i$ is right-censored ($\Delta_i=0$) or not ($\Delta_i=1$).
Here, it is assumed that each individual can experience one out of $J$ competing events and that the event type of the $i$-th individual at $T_i$ is denoted by $\epsilon_i \in \{1,\hdots, J\}$. In accordance with \citet{Fine1999}, our interest is in modeling the cumulative incidence function $F_1(t) = P(T \le t, \epsilon = 1)$ of a type~1 event conditional on covariates, taking into account that there are $J-1$ competing events and also the censoring event ($\Delta_i=0$).

For given values of a set of time-constant covariates $\bs{x}_i=(x_{i1},\hdots,x_{ip})^\top$, the discrete subdistribution hazard function for a type~1 event is defined by 
\begin{align}
\lambda_1(t|\bs{x}_i)\, &= \, P\left(T_i=t,\epsilon_i=1| (T_i\geq t) \cup (T_i\leq t-1,\epsilon_i \neq 1), \bs{x}_i \right) \\ \label{eq:hazard}
&=\, P(\vartheta_i=t|\vartheta_i\geq t,\bs{x}_i) \, ,
\end{align}
where (\ref{eq:hazard}) is the discrete hazard function of the subdistribution time 
\begin{align}
\vartheta_i :=\begin{cases}T_i, \text{ if } \epsilon_i=1 \, , \\ \infty, \text{ if } \epsilon_i\neq 1\, ,
\end{cases}
\end{align}
cf.\@ \citet{berger2018sub}. The subdistribution time $\vartheta_i$ measures the time to the occurrence of a type~1 event first and is not finite if $\varepsilon_i \neq 1$ (as a type 1 event will never be the first event as soon as a competing event has occurred). 
The discrete subdistribution hazard is linked to the cumulative incidence function by 
\begin{align}\label{eq:cif}
F_1 (t|\bs{x}_i) = 1 - \prod_{s=1}^{t}{(1-\lambda_1(s|\bs{x}_i))} = 1 - S_1(t|\bs{x}_i ) \, , 
\end{align} 
where $S_1(t|\bs{x}_i)=P(\vartheta_i>t|\bs{x}_i)$ is the discrete survival function for a type~1 event. Thus, a regression model for the discrete subdistribution hazard $\lambda_1$ has a direct interpretation in terms of the cumulative incidence function $F_1$.

A class of regression models that relate the discrete subdistribution hazard function \eqref{eq:hazard} to the covariates $\bs{x}_i$ was proposed by \citet{berger2018sub}. It is defined by 
\begin{align} \label{eq:mod1}
\lambda_1(t|\bs{x}_i)=h(\eta_1(t,\bs{x}_i))\,,
\end{align}
where $h(\cdot)$ is a strictly monotone increasing distribution function. In line with classical hazard models for discrete event times (e.g.\@\citealt{Tutz2016}), it is assumed that the predictor function 
\begin{align} \label{eq:modelparams}
\eta_1(t,\bs{x}_i) = \gamma_{0t}+\bs{x}_i^\top\bs{\gamma}\,,
\end{align}
is composed of a set of time-varying intercepts $\gamma_{01},\hdots,\gamma_{0,k-1}$, referred to as {\em baseline coefficients}, and a linear function of the covariates with coefficients $\bs{\gamma} \in \mathbb{R}^p$ that do not depend on $t$. As in generalized additive models it is also possible to extend $\eta_1(t,\bs{x}_i)$ by interactions and smooth (possibly non-linear) functions. A~popular choice of $h(\cdot)$ is the inverse complementary log-log function, which yields the \textit{Gompertz model} $\lambda_1(t,\bs{x}_i)=1-\exp(-\exp(\eta_1(t,\bs{x}_i)))$ that is equivalent to the original subdistribution hazard model by \citet{Fine1999} for continuous time-to-event data. 

\section{Model fitting} \label{sec:modelFitting}

In \citet{berger2018sub} we showed that estimates of the model parameters in~\eqref{eq:modelparams} can be derived using estimation techniques for weighted binary regression. This approach is based on the observation that with i.i.d.\@ data $(\tilde{T}_i,\Delta_i,\epsilon_i, \bs{x}_i)$, $i=1,\hdots,n$, the log-likelihood of model \eqref{eq:mod1} can be expressed as 
\begin{align}\label{eq:logL}
\ell \, = \, \sum_{i=1}^{n}\sum_{t=1}^{k-1}{w_{it}\big\{y_{it} \log(\lambda_1(t|\bs{x}_i))+(1-y_{it}) \log(1-\lambda_1(t|\bs{x}_i))\big\}}\,
\end{align}
with binary outcome values 
\begin{align}\label{eq:responses}
(y_{i1},\hdots,y_{i,\tilde{T}_i},\hdots,y_{i,k-1})=\begin{cases}(0,\hdots,0,1,0,\hdots,0), \;\text{ if }\; \Delta_i\epsilon_i=1\, ,\\(0,\hdots,0,0,0,\hdots,0), \;\text{ if }\; \Delta_i\epsilon_i \neq 1\,.\end{cases}
\end{align}
For uncensored individuals that experience a type 1 event first $(\Delta_i\varepsilon_i=1)$ and for censored individuals $(\Delta_i\varepsilon_i=0)$ the weights $w_{it}$ are defined by 
\begin{align}\label{eq:w1}
w_{it}:=\text{I}(t\leq \tilde{T}_i)\, ,
\end{align}
whereas for uncensored individuals experiencing a competing event first $(\Delta_i\varepsilon_i>1)$ they are defined by 
\begin{align}\label{eq:w2}
w_{it}:=\begin{cases}1,\;&\text{ if }\; t\leq\tilde{T}_i\,,\\\frac{\hat{G}(t-1)}{\hat{G}(\tilde{T}_i -1)},\;&\text{ if }\; \tilde{T_i} < t \le k-1 \, .\end{cases}
\end{align}
The function $\hat{G}(t)$ in \eqref{eq:w2} is an estimate of the censoring survival function $G(t)= {P}(C_i>t)$, implying that the weights in \eqref{eq:w1} and \eqref{eq:w2} equal estimates of the individual-specific conditional probabilities of being (still) at risk for a type 1 event at time $t$. As shown in \citet{berger2018sub}, maximization of the log-likelihood \eqref{eq:logL} yields consistent and asymptotically normal estimators of the parameters $\gamma_{0t}$ and $\bs{\gamma}$. In Sections~\ref{sec:calplot} and~\ref{sec:logistic} we will show that the weights defined in \eqref{eq:w1} and \eqref{eq:w2} also play a key role in the calibration assessment of discrete-time subdistribution hazard models.

\section{Calibration plot} \label{sec:calplot}

In the following we will assume that an i.i.d.\@ training sample $(\tilde{T}_i,\Delta_i, \epsilon_i, \bs{x}_i),\,i=1,\hdots,n$ has been used to fit
a statistical model that can be used to predict the indivual-specific subdistribution hazards $\lambda_1 (t| \bs{x})$ in some study population.
We will further assume that the calibration of the fitted model is assessed by means of an independent i.i.d.\@ validation sample with $N$ individuals $(\tilde{T}_m,\Delta_m,\epsilon_m, \bs{x}_m),\,m=1,\hdots,N$.
The starting point of our considerations is the calibration plot proposed in \citet{Berger2018}, which applies to discrete hazard models with only a single type of event $(J=1)$. Note that both the specification of the subdistribution hazard model and the definition of its log-likelihood function remain valid in this case, as the scenario without competing events $(J=1)$ is a special case of equations \eqref{eq:mod1} and \eqref{eq:logL}. The idea underlying the method by \citet{Berger2018} is to split the test data into $G$ subsets $D_g,\,g=1,\hdots,G$, defined by the percentiles of the {\em predicted} hazards $\hat{\lambda}_1(t|\bs{x}_m) = \hat{P}(T_m = t | T_m \ge t, \bs{x}_m) = \hat{P}(y_{mt} = 1 | \bs{x}_m), \,t=1,\hdots,\tilde{T}_m,\,m=1,\hdots,N$, which are obtained from the fitted binary model in~\eqref{eq:mod1}. Following the approach by \cite{hosmer2013} for assessing the calibration of binary regression models, the average predicted hazards in the $G$ groups are subsequently plotted against the \textit{empirical} hazards, which are given by the group-wise relative frequencies of outcome values with $y_{tm} = 1$. A well calibrated model is indicated by a set of points that is close to the 45-degree line.

More formally, the predicted and empirical hazard estimates considered in \citet{Berger2018} can be written as
\begin{align}\label{eq:calplot2}
&\overline{\hat{\lambda}}_{1g}=\frac{1}{\sum\limits_{m,t: \hat{\lambda}_1(t|\bs{x}_m) \in D_g}{w_{mt}}}\;\sum_{m,t: \hat{\lambda}_1(t|\bs{x}_m) \in D_g}{\hat{\lambda}_1(t|\bs{x}_m)\,w_{mt}}\,,
\nonumber\\
\text{and}\quad
&\overline{y}_g=\frac{1}{\sum\limits_{m,t: \hat{\lambda}_1(t|\bs{x}_m) \in D_g}{w_{mt}}}\;\sum_{m,t: \hat{\lambda}_1(t|\bs{x}_m) \in D_g}{y_{mt}\,w_{mt}}\,,\nonumber\\
&t=1,\hdots,k-1,\;m=1,\hdots,N\,,
\end{align}
respectively, where $w_{mt}:=I(t\leq\tilde{T}_m) \in \{0,1\}$ indicates whether individual $m$ is at risk for a type 1 event at time point $t,\,t=1,\hdots,k-1$, or not. Note that the definition of $w_{mt}$ in \eqref{eq:calplot2} is exactly the same as the definition of the weight $w_{it}$ in~\eqref{eq:w1}. Also note that $\sum_{m,t: \hat{\lambda}_1(t|\bs{x}_m) \in D_g}{w_{mt}}=|D_g|$, as only the values $\hat{\lambda}_{1}(t| \bs{x}_m)$ with $w_{mt}=1$ are used for defining the groups $D_1,\ldots, D_g$. In a well calibrated hazard model, the values $\overline{\hat{\lambda}}_{1g}$ should be close to their counterparts $\overline{y}_g$. 

Now consider the scenario where, in addition to the type 1 event of interest, competing events of type $2,\hdots,J$ may be observed. In this case, $\lambda_1$ becomes the subdistribution hazard of a type 1 event, as defined in \eqref{eq:hazard}. To obtain a calibration plot for a fitted subdistribution hazard model, we propose to define the quantities $\bar{\hat{\lambda}}_{1g}$ and $\bar{y}_{g}$ analogously to the single-event scenario considered in~\eqref{eq:calplot2}. Unlike in the scenario with $J=1$, however, the definition of the terms $w_{mt}$ is not straightforward: The problem is that individuals experiencing a competing event first continue to be at risk beyond $\tilde{T}_m$ until they experience the censoring event. Hence, as the censoring times $C_m$ are unobserved if $C_m>\tilde{T}_m$, it usually cannot be determined whether these individuals are still at risk at $t>\tilde{T}_m$. In accordance with \citet{berger2018sub}, we therefore propose to predict the probability of each individual $m=1,\hdots,N$, of being at risk for a type 1 event at time $t$ and to set the terms $w_{mt}$ equal to the predicted probabilities.

More specifically, the proposed strategy comprises the following steps:
\begin{itemize}
\item[(i)] Sort the predicted subdistribution hazards $\hat{\lambda}(t | \bs{x}_m)$, $t=1,\ldots , k-1$, $m=1, \ldots , N$, obtained from the fitted subdistribution hazard model and form groups $D_1,\ldots , D_G$ defined by the percentiles of $\hat{\lambda}(t | \bs{x}_m)$.
\item[(ii)] Compute the weights $w_{mt}$ using the formulas in \eqref{eq:w1} and \eqref{eq:w2}, where $\hat{G}(\cdot)$ is estimated from the learning sample $i=1,\hdots,n$.
\item[(iii)] Compute $\bar{\hat{\lambda}}_{1g}$ and $\bar{y}_{g}$ as in \eqref{eq:calplot2} using the weights obtained in step (ii). Note that by definition, $\sum_{m,t: \hat{\lambda}_1(t|\bs{x}_m) \in D_g}{w_{mt}}\leq |D_g|$.
\item[(iv)] Plot $\bar{\hat{\lambda}}_{1g}$ against $\bar{y}_{g}$.
\item[(v)] Assess the calibration of the fitted subdistribution hazard model by inspecting the plot generated in step (iv). A well calibrated model is indicated by a set of points that is close to the 45-degree line.
\end{itemize}

\section{Recalibration model} \label{sec:logistic}

In addition to the graphical checks presented in Section \ref{sec:calplot}, we propose a recalibration approach for discrete subdistribution hazard models originating from the method by \citet{cox1958}. The idea of this method, which was originally developed for assessing the calibration of binary regression models, is to fit a logistic regression model to the test data in order to investigate the agreement between a set of predicted probabilities and the respective values of the binary outcome variable.

Based on the binary representation of the subdistribution hazard model in \eqref{eq:mod1} and \eqref{eq:logL}, we propose to adapt the recalibration framework by \citet{cox1958} as follows: Assuming that calibration assessments are again based on a validation sample $(\tilde{T}_m , \Delta_m , \epsilon_m , \bs{x}_m)$, $m=1,\hdots,N$, we propose to fit a logistic regression model of the form
\begin{align}\label{eq:recal_mod}
\log\left(\frac{\lambda_1(t|\bs{x}_m)}{1-\lambda_1(t|\bs{x}_m)}\right)=&\,\eta_{\text{rc}}(t|\bs{x}_m)= a+b\,\log\left(\frac{\hat{\lambda}_1(t|\bs{x}_m)}{1-\hat{\lambda}_1(t|\bs{x}_m)}\right)\,,\nonumber \\
&\,t=1,\hdots,k-1,\;m=1,\hdots,N\,,
\end{align}
where $\hat{\lambda}_1(t|\bs{x}_m)$ are the predicted hazards defined in Section \ref{sec:calplot}. The intercept $a$ in model \eqref{eq:recal_mod} measures ``calibration-in-the-large'', i.e.\@ it indicates whether the predicted hazards are systematically too low ($a>0$) or too high ($a<0$). Analogously, the slope~$b$ measures ``refinement'', which indicates that the predicted hazards either do not show enough variation ($b>1$), vary too much ($0<b<1$), or show the wrong general direction ($b<0$, \citealt{miller1993}).

To assess the fit of the predicted hazards we propose to follow the suggestions by \citet{miller1993} and to conduct recalibration tests on the following null hypotheses: (i)~$H_0$:~$a=0,\,b=1$, which refers to an overall test for calibration, (ii)~$H_0$:~$a=0\,|\,b=1$, to test for calibration-in-the-large given appropriate refinement, and (iii)~$H_0$:~$b=1\,|\,a$, to test refinement given corrected calibration-in-the-large. 

Because the predicted hazards $\hat{\lambda}_1$ are derived from a subdistribution hazard model that was fitted using weighted maximum likelihood estimation, we propose to fit the recalibration model in \eqref{eq:recal_mod} by optimizing a weighted binary log-likelihood of the form 
\begin{align}\label{eq:logLik} 
\ell_{\text{rc}} \, = \, \sum_{m=1}^{N}\sum_{t=1}^{k-1}{w_{mt}\big\{y_{mt} \log(\pi_1(t|\bs{x}_m))+(1-y_{mt}) \log(1-\pi_1(t|\bs{x}_m))\big\}}\,,
\end{align}
where the probabilities $\pi_1 (t| \bs{x}_m)$ are given by $\pi_1 (t| \bs{x}_m) = \exp (\eta_{\text{rc}}(t|\bs{x}_m )) / (1 + \exp ( \eta_{\text{rc}}(t| \bs{x}_m))$. The binary outcome values $y_{mt}$ and the weights $w_{mt}$ are defined in the same way as in Section \ref{sec:calplot}.  Note that $\hat{G}(\cdot)$ is again estimated from the learning sample $i=1,\hdots,n$. In case $a=0$ (referring to the tests in (i) and (ii) above), the log-likelihood \eqref{eq:logLik} can be written as 
\begin{align}\label{eq:logLik2}
\ell_{\text{rc}}\, = \, &b\,\sum_{m=1}^{N}\sum_{t=1}^{k-1}{w_{mt} y_{mt} \log\left(\hat{\lambda}_1(t|\bs{x}_m)\right) } \nonumber \\
\, &+ b\,\sum_{m=1}^{N}\sum_{t=1}^{k-1}{w_{mt} (1-y_{mt}) \log\left(1-\hat{\lambda}_1(t|\bs{x}_m)\right)} \nonumber \\ \, &- \sum_{m=1}^{N}\sum_{t=1}^{k-1}{w_{mt} \log \left(\hat{\lambda}_1(t|\bs{x}_m)^b+(1-\hat{\lambda}_1(t|\bs{x}_m))^b\right)}\,,
\end{align}
which corresponds to the weighted log-likelihood in Equation~(7) of \citet{cox1958}. The derivation of \eqref{eq:logLik2} is given in Appendix~\ref{app:A}. It follows that hypotheses (i) to (iii) can be examined using likelihood-ratio test statistics that asymptotically ($N\to\infty$) follow $\chi^2$-distributions with one (hypotheses~ii and~iii) and two (hypothesis~i) degrees of freedom.  

\section{Numerical Experiments} \label{sec:sim} 

In this section we present the results of numerical experiments to evaluate the proposed calibration measures under known conditions. The main focus of the study was on measuring the performance in scenarios with different rates of type~1 events, different levels of censoring and a varying number of discrete time points. 

\subsection{Experimental Design} \label{sec:sim_design}

In order to generate data from a given subdistribution hazard model for type 1 events, we used a scheme adopted from \citet{Fine1999}. This procedure is also described in \citet{beyersmann2011}, where it was termed ``indirect simulation''. In all simulation scenarios we considered data with two competing events, $\epsilon_i \in \{1,2\}$, that was generated under the model specification of proportional subdistribution hazards. More specifically, our discrete subdistribution hazard model was based on the discretization of the continuous model \vspace{-.2cm}
\begin{equation}\label{eq:sim_F1}
F_1(t|\bs{x}_i)=P(T_{cont,i}\leq t,\epsilon_i=1|\bs{x}_i)=1-(1-q+q\exp(-t))^{\exp(\bs{x}_i^\top\bs{\gamma})}\,, 
\vspace{-.2cm}
\end{equation}
where $T_{cont,i} \in \mathbb{R}^+$ denotes the continuous time span of individual $i$ and $\bs{\gamma}=(\gamma_{1},\hdots,\gamma_{p})^\top$ is a set of regression coefficients. The parameter $q\in(0,1)$ affected the probability of a type 1 event which, according to \eqref{eq:sim_F1}, was given by  $\pi_{i1} := P(\varepsilon_i=1|\bs{x}_i)=1-(1-q)^{\exp(\bs{x}_i^\top\bs{\gamma})}$. By definition, high values of $q$ resulted in high probabilities of $\pi_{i1}$, and vice versa. The probability of a competing event was given by $\pi_{i2}:=P(\epsilon_i=2|\bs{x}_i)=1-\pi_{i1}=(1-q)^{\exp(\bs{x}_i^\top\bs{\gamma})}$.

Continuous time spans for type 2 events were drawn from the exponential model
\begin{align*}
T_{cont,i}|\epsilon_i=2,\bs{x}_i \sim \text{Exp}(\xi_2 =\exp(\bs{x}_i^\top\bs{\beta}))\,,
\end{align*}
where $\bs{\beta}=(\beta_1,\hdots,\beta_p)^\top$ denotes a set of regression coefficients linking the rate parameter $\xi_2$ with the values of the covariates $\bs{x}$. 

In order to obtain discrete event times, we generated data according to the indirect simulation scheme described above and grouped the resulting continuous event times into various numbers of categories ($k=5, 10, 15$). The latter were defined by the quantiles of the continuous event times, which were pre-estimated from an independent sample with 1,000,000 observations. Consequently, the same interval boundaries were used in each simulation run. Censoring times were generated from a discrete distribution with probability density function $P(C_{disc,i}=t)=b^{k-t+1}/\sum_{s=1}^{k}b^{s}$, $t=1,\hdots,k$,
where the percentage of censored observations was controlled by the parameter $b\in \mathbb{R}^+$.

We considered two standard normally distributed covariates $x_{i1},x_{i2} \sim N(0,1)$ and two binary covariates $x_{i3},x_{i4} \sim B(1,0.5)$. All covariates were independent, and the true regression coefficients were set to $\bs{\gamma}=(0.4, -0.4, 0.2, -0.2)^\top$ and $\bs{\beta}=(-0.4, 0.4, -0.2, 0.2)^\top$, cf.\@ \citet{Fine1999}. We specified three different censoring rates, denoted by {\it weak}, {\it medium} and {\it strong}, where the degree of censoring was controlled by the parameter $b$ of the censoring distribution. More specifically, we used the values $b=0.85$ (weak), $b=1$ (medium) and $b=1.25$ (strong), resulting in the censoring rates shown in Figure \ref{fig:sim_design} in Appendix \ref{app:B}. We also considered three different probabilities of a type~1 event, specifying $q\in\{0.2,0.4,0.8\}$. In total, this resulted in $3\times 3 \times 3 = 27$ different scenarios. All scenarios were analyzed using 100 replications with $5000$ independently drawn observations each that were equally split into a learning sample and a validation sample ($n=2500$ and $N=2500$). For estimation we used the inverse complementary log-log function (Gompertz model), which defines the same values of~$\bs{\gamma}$ as the Fine \& Gray proportional subdistribution hazards model in continuous time.

Figure \ref{fig:sim_design} in Appendix \ref{app:B} illustrates the relative frequencies of observed events for the nine scenarios with $k = 5$.  It is seen that the rates of observed type 1 events increased with increasing value of $q$ and that censoring rates increased with increasing value of $b$. For constant $q$ and varying~$b$, the ratio of observed type 1 and type 2 events remained approximately the same. For $q=0.2$ and $q=0.4$ we observed more events of type 2 than of type~1, and for $q=0.8$ there were more events of type 1 than of type~2. For the scenarios with $k = 10$ and $k = 15$ the observed relative frequencies were almost the same and are thus not shown.

\begin{figure}[!t]
\centering
\includegraphics[width=1\textwidth]{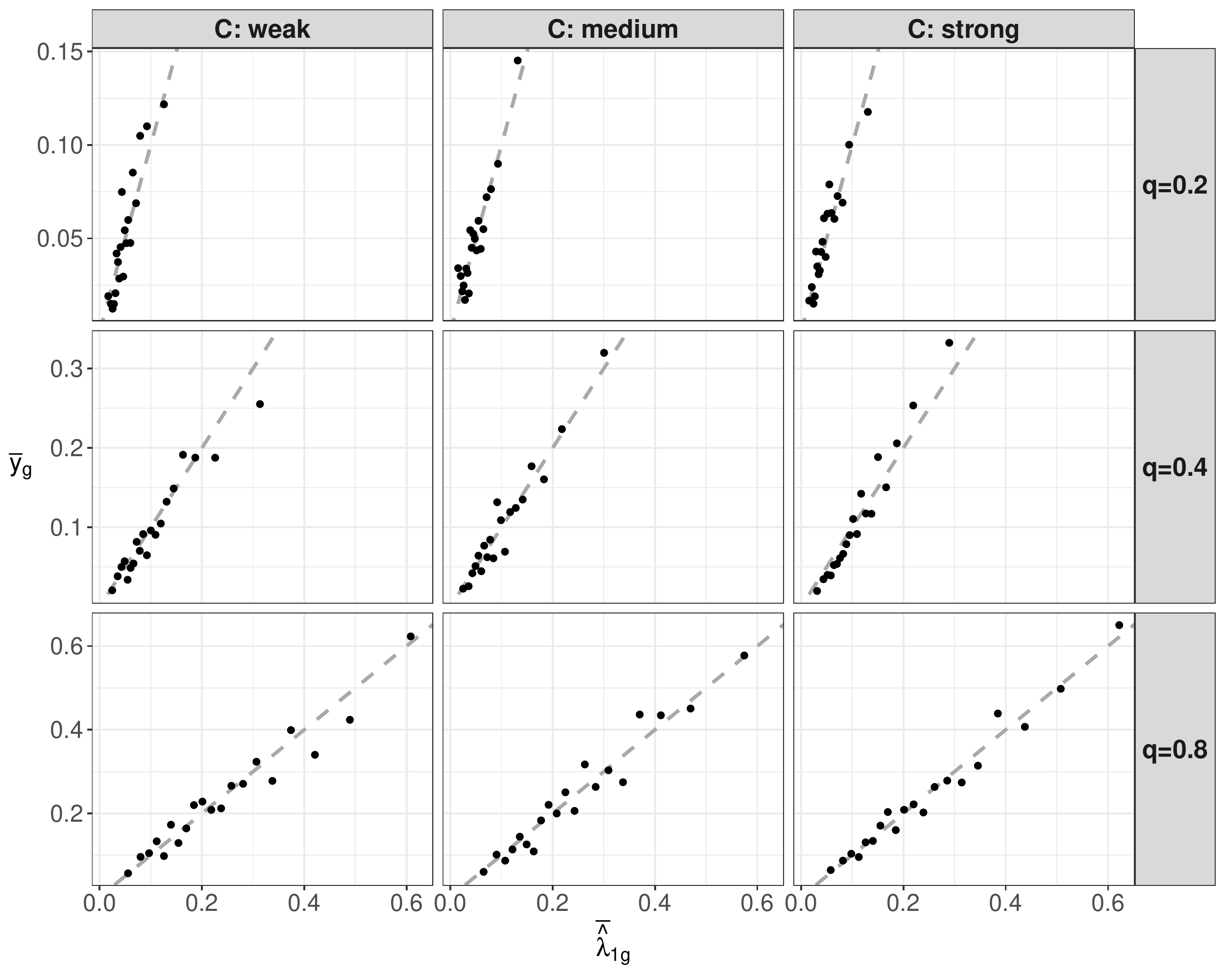}
\caption{Results of the simulation study. Calibration plots of one randomly chosen replication in each simulation scenario using $G=20$ subsets $(k=5)$. The 45-degree lines (dashed) indicate perfect calibration (C = degree of censoring).}
\label{fig:sim_plots}
\end{figure}

\begin{figure}[!t]
\centering
\includegraphics[width=1\textwidth]{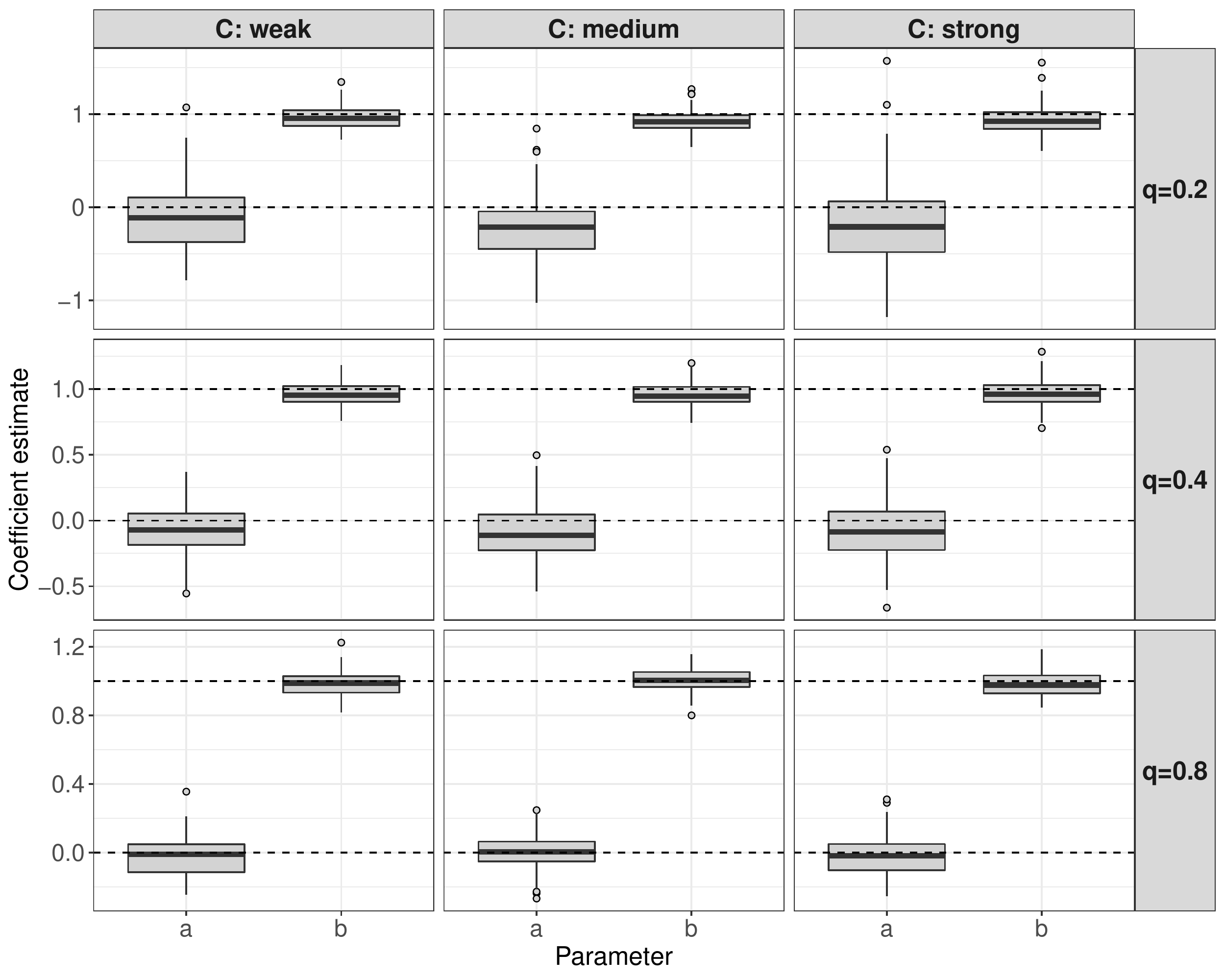}
\caption{Results of the simulation study. The boxplots visualize the estimates of the calibration intercepts $a$ and calibration slopes $b$ that were obtained from fitting the logistic recalibration model $(k=5)$.}
\label{fig:sim_coefs}
\end{figure}

\begin{figure}[!t]
\centering
\includegraphics[width=1\textwidth]{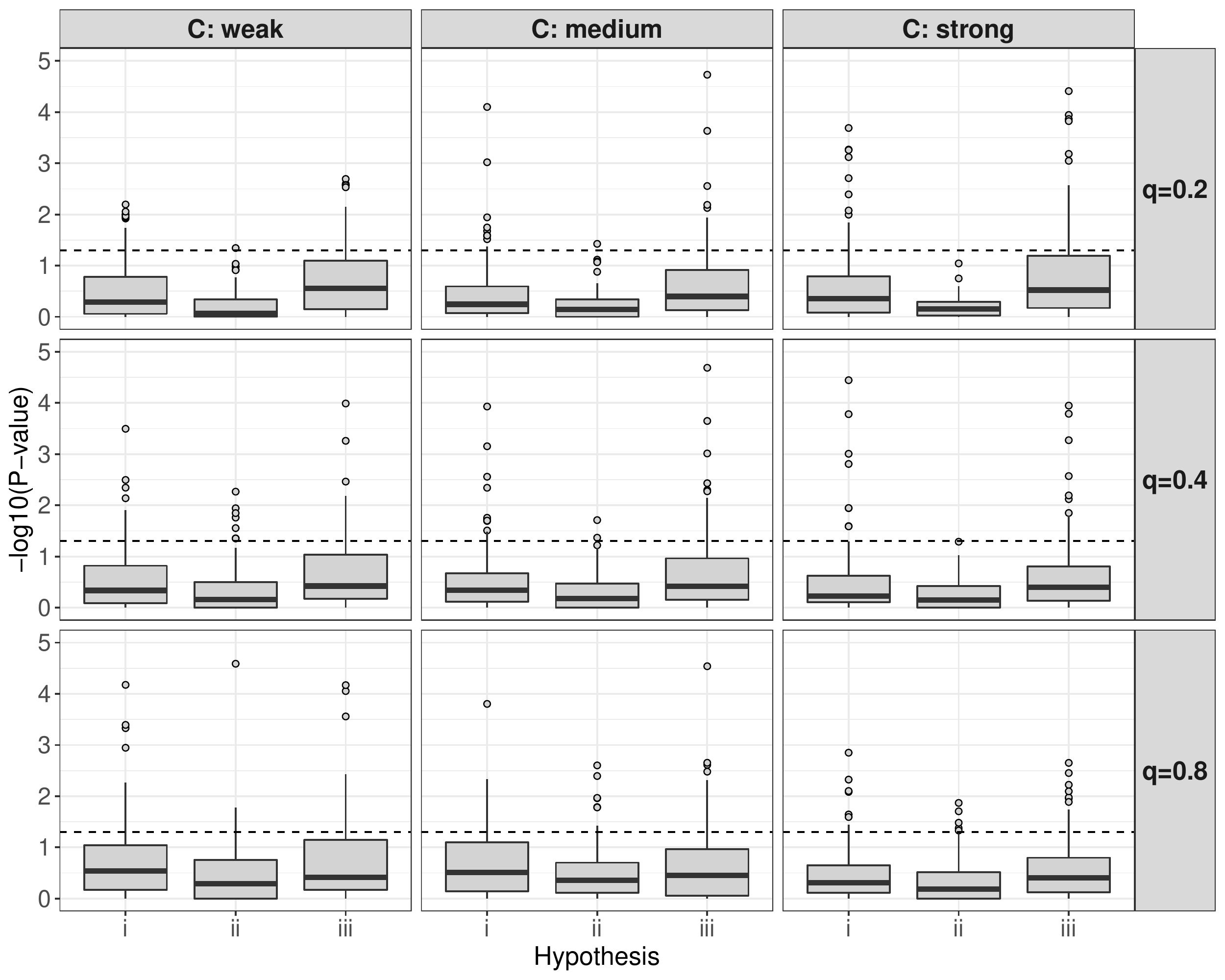}
\caption{Results of the simulation study. The boxplots visualize the negative log10-transformed $p$-values obtained from the recalibration tests~$(k=5)$. The dashed lines correspond to a $p$-value of $0.05$. A value above the dashed line indicates a significant result at the $5\%$ type 1 error level.}
\label{fig:sim_pvalues}
\end{figure}

\subsection{Results} \label{sec:sim_res}

The calibration plots for one randomly chosen replication of the nine simulation scenarios with $k=5$ are presented in Figure \ref{fig:sim_plots}. The plots show that the empirical hazards $\overline{y}_g$ and the average predicted hazards $\overline{\hat{\lambda}}_{1g}$ strongly coincided, regardless of the degree of censoring and the rate of type 1 events.
This result illustrates that the calibration plot defined in Section \ref{sec:calplot} is a strong tool for the graphical assessment of a correctly specified discrete subdistribution model.
Figure \ref{fig:sim_plots} further shows that modeling the subdistribution hazard $\lambda_1$ also worked well in the presence of only a relatively small number of type 1 events (for $q=0.2$ only about $10\%$ type~1 events were observed). Note that the $y$-axis limits differ across the rows, which is the reason why the points are not spread over the whole plots for $q=0.2$ and $q=0.4$. When fitting the logistic recalibration model, for example to the data set with strong censoring and $q=0.2$ (upper right panel of Figure \ref{fig:sim_plots}), we obtained the estimates $\hat{a}=-0.084$ and $\hat{b}=0.957$, which were rather close to values zero and one in case of perfect calibration. 

Exemplary calibration plots for the scenarios with $k=10$ and $k=15$ are presented in Figure \ref{fig:sim_plots10} and Figure \ref{fig:sim_plots15} in Appendix \ref{app:B}. Again, the plots suggested nearly perfect calibration, with the exception of the scenarios with $k=15$ and strong censoring, where the variation and thus the deviation from the 45-degree lines was more apparent.  

The estimates of the calibration parameters $a$ and $b$ for all scenarios with $k=5$ are shown in Figure \ref{fig:sim_coefs}. It is seen from the boxplots that on average the estimates were very close to values zero and one. In particular, for $q=0.8$ (lower panel) the results of the recalibration model correctly indicated nearly perfect calibration. It is also seen that the variance of the estimates of the intercept $a$ increased with decreasing rate of type 1 events. In contrast, the degree of censoring had only a small impact on the variance of the estimates. Figure \ref{fig:sim_pvalues} presents the corresponding $p$-values when conducting the recalibration tests (i) -- (iii) specified in Section~\ref{sec:logistic}. Throughout all scenarios the null hypotheses were kept in almost all replications (at the $5\%$ type 1 error level). In particular the tests for calibration-in-the large given appropriate refinement (hypothesis ii) yielded very large $p$-values (corresponding to small negative log10-transformed values). For example, in the scenario with $q=0.2$ and strong censoring this hypothesis was never rejected at the $5\%$ type 1 error level. Overall, the results in Figure \ref{fig:sim_coefs} and \ref{fig:sim_pvalues} illustrate that the proposed logistic recalibration model properly assessed the calibration of the fitted subdistribution hazard models, even in the case of strong censoring and a small rate of type 1 events. 

The parameter estimates ($\hat{a}$ and $\hat{b}$) and the $p$-values for the scenarios with $k=10$ and $k=15$ are given in Figures \ref{fig:sim_coefs10} to \ref{fig:sim_pvalues15} in Appendix \ref{app:B}. These results largely confirmed the previous findings for $k=5$. Although the estimated calibration parameters deviated more strongly from zero and one, the associated null hypotheses were still kept at the $5\%$ type 1 error level. The only exceptional case was the scenario with $k=15$, strong censoring and $q=0.2$ (upper right panel of Figure \ref{fig:sim_pvalues15}), where about half of the null hypotheses (i) and (iii) were rejected. These results, which were clearly related to the number of time intervals can be explained by the fact that very few type 1 events were observed at later points in time when $k$ was increased. For example, with $k=15$, strong censoring and $q=0.2$, less than 4 type 1 events occurred at time points $t>10$ in most of the learning and validation samples.

\section{Nosocomial pneumonia infection in intensive care units} \label{sec:app}

To illustrate the use of the proposed calibration measures, we analyzed a data set on the development of pneumonia, which is a common nosocomial, i.e.\@ hospital-acquired infection in intensive care units (ICUs). This data set was also considered before by \citet{beyersmann2006}, \citet{wolkewitz2008}, \citet{ berger2018sub} and other authors. As nosocomial pneumonia (NP) has a strong impact on the mortality of patients in ICUs, it is of high interest to determine risk factors for the development of the disease. 

The data were collected for a prospective cohort study at five ICUs in one university hospital, lasting 18 months from February 2000 to July 2001 and comprises $n=$1,876 patients with a duration of ICU stay of at least two days.  The outcome of interest was the time to NP infection. Other possible events that were competing with the onset of NP (being the event of interest) were \textit{death} and \textit{discharge from hospital alive}. Due to the study design, the observed event times were discrete, as they were measured on a daily basis. \citet{berger2018sub} analyzed the data over a period of 60 days, resulting in 61 possible event times $t=1,2,\hdots,61$, where $t=k=61$ refered to all individuals with event times $\geq 61$ days. At the observed times, each patient either acquired the NP infection ($n=158$), was released from  hospital (alive) or died ($n=$1,695), or was administratively censored~($n=23$).   
Descriptive summary statistics of the baseline risk factors considered in the analysis were presented in Table $1$ of \citet{wolkewitz2008}. In addition to the age of the patients (centered at 60 years), the gender of the patients, and the simplified acute physiology score (SAPS)~II, there were eleven binary risk factors characterizing the patients and their hospital stay. The binary variables either referred to the time of ICU admission ({\it on admission}) or the time prior to ICU admission ({\it before admission}).

To analyze the effects on the time to NP infection, we fitted the discrete subdistribution hazard model used in \citet{berger2018sub}. This model incorporates the baseline risk factors and a set of smooth baseline coefficients represented by cubic P-splines with a second-order difference penalty (fitted using the R package \textbf{mgcv}). It was referred to as Model~2 in \citealp{berger2018sub}. To assess the calibration of the model, we conducted a benchmark experiment that was based on 100 random partitions of the data. Each partition consisted of a learning sample of size $n = 1,500\,(80\%)$ and a validation sample of size $N = 376\,(20\%)$.

\begin{figure}[!t]
\centering
\includegraphics[width=0.48\textwidth]{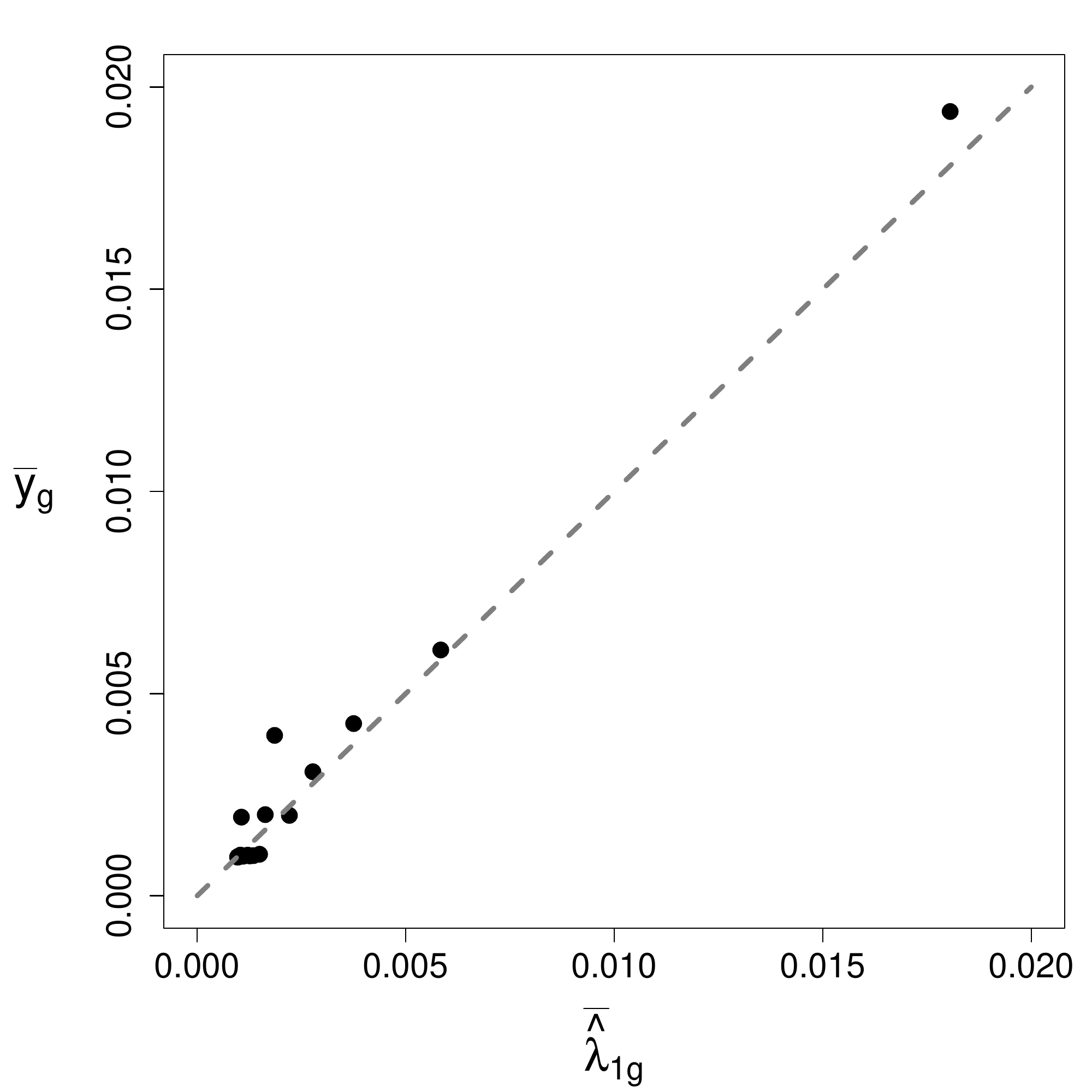}
\caption{Analysis of the NP infection data. Calibration plot obtained for a randomly chosen partition of the data into learning and validation sample using $G=20$ subsets. The 45-degree line (dashed line) indicates perfect calibration.}
\label{fig:app_plot}
\end{figure}

\begin{figure}[!t]
\centering
\includegraphics[width=0.4\textwidth]{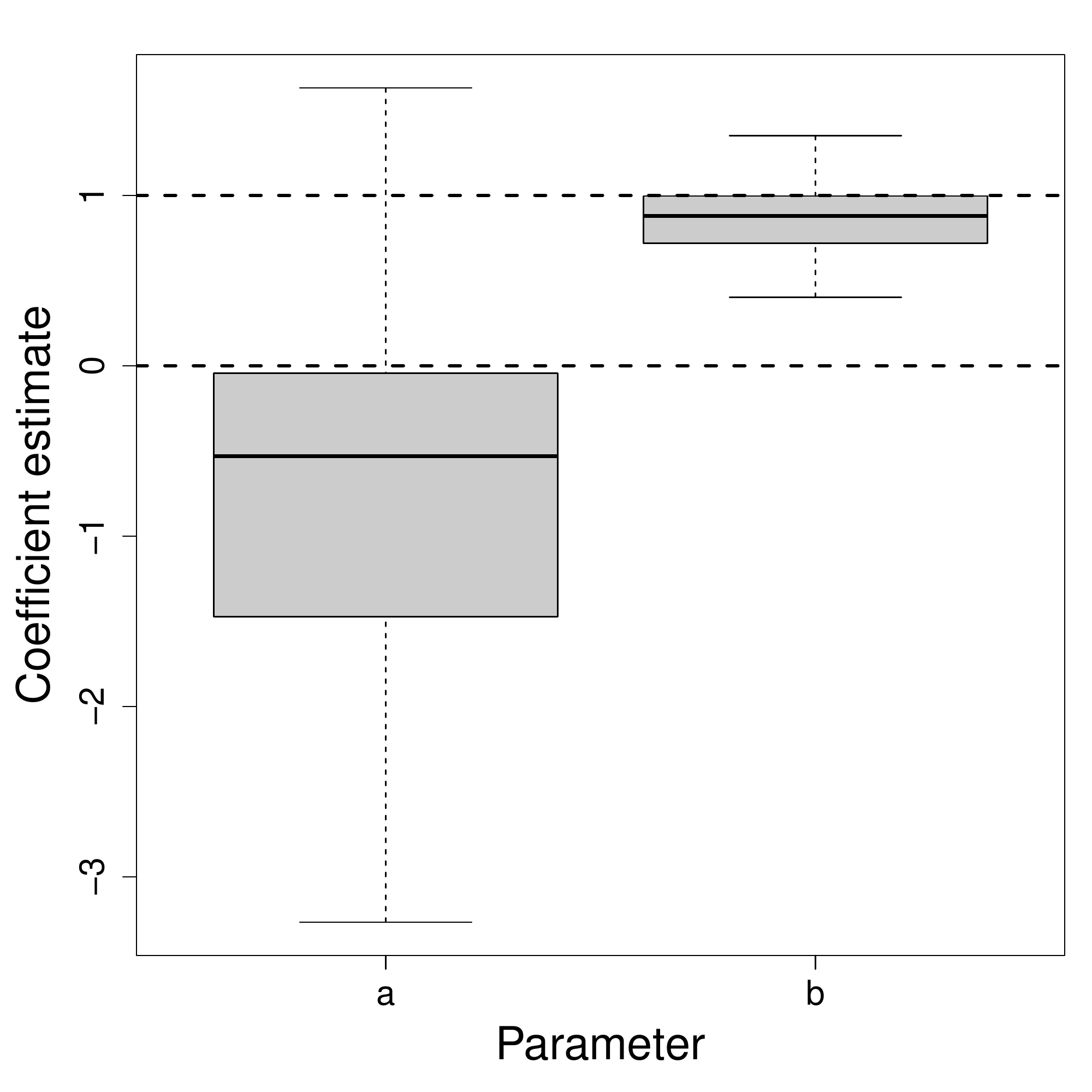}
\includegraphics[width=0.4\textwidth]{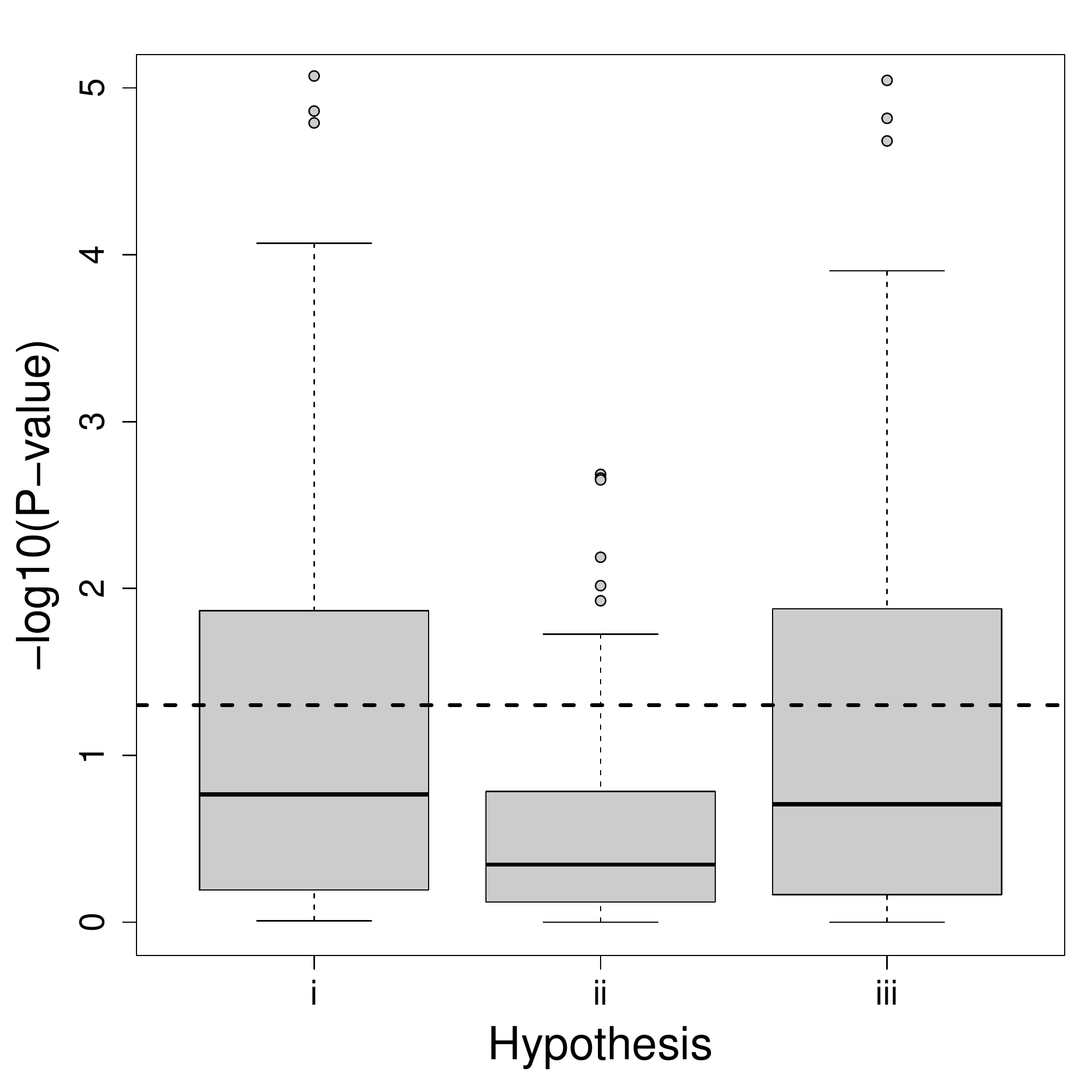}
\caption{Analysis of the NP infection data. Estimates of the calibration intercepts $a$ and calibration slopes $b$ (left) and negative log10-transformed $p$-values obtained from the recalibration tests (right) obtained for $100$ partitions of the data into learning and validation sample.}
\label{fig:app_coefs}
\end{figure}

Figure \ref{fig:app_plot} presents the calibration plot of the model that was obtained for one randomly chosen  partition of the data. It is seen that apart from the two subsets defined by the largest percentiles, the empirical hazards $\overline{y}_g$ and the average predicted hazards $\overline{\hat{\lambda}}_{1g}$ were rather small~\mbox{$(< 0.005)$}. Furthermore, the plot showed strong coincidence between~$\overline{y}_g$ and $\overline{\hat{\lambda}}_{1g}$, indicating satisfactory calibration of the fitted model. Boxplots of the estimated calibration parameters $a$ and $b$ and the $p$-values when performing the associated recalibration tests  are shown in Figure~\ref{fig:app_coefs}. The estimates related to the calibration plot in Figure~\ref{fig:app_plot} were $\hat{a}=0.039$ and $\hat{b}=0.984$ with $p$-values $0.809$~(hypothesis i), $0.518$~(hypothesis ii) and $0.935$~(hypothesis~iii). The average estimates of $a$ and $b$ (left panel of Figure \ref{fig:app_coefs}) indicated that the predicted hazards tended to be systematically to high $(a<0)$ and that they varied a little to much $(0 < b < 1)$. Importantly, this trend was also seen in the simulation scenarios with weak censoring and $q=0.2$ (cf. Figure \ref{fig:sim_coefs} and Figure \ref{fig:sim_coefs10} and \ref{fig:sim_coefs15} in Appendix \ref{app:B}), which is the setting that is most comparable to the characteristics of the NP infection data. Also note that the number of observed type 1 events in the data is smaller than 3 at time points $t>20$. According to the recalibration tests  (right panel of Figure \ref{fig:app_coefs}), the deviations of the calibration parameters from zero and one were not substantial, as the majority of the null hypotheses on proper calibration-in-the-large and refinement were not rejected (at the 5\% type 1 error level). This result is again in line with the findings in the simulation study and also indicated a good calibration of the discrete subdistribution model used in \citet{berger2018sub}.

\section{Concluding remarks} \label{sec:conc}

Discrete time-to-event models have gained considerable popularity during the past years \citep{Tutz2016, lee2018}. Therefore, methodology for the proper validation of their generalization performance is increasingly necessary. In this regard, the methods presented here constitute a new set of tools to assess the calibration of discrete subdistribution hazard models for competing risks analysis. They consist of a calibration plot for graphical assessments as well as a recalibration model including tests on calibration-in-the-large and refinement. Both methods are well connected to analogous approaches for binary regression \citep{hosmer2013, miller1993}. In the single-event scenario, the graphical tool presented here naturally reduces to the calibration plot proposed in \citet{Berger2018}. 

Unlike \citet{heyard2019}, who proposed tools to assess the calibration of cause-specific hazard models, we considered the subdistribution framework originally proposed by \citet{Fine1999} for competing risks data in continuous time. In contrast to cause-specific hazard modeling, this approach has the advantage that only one model needs to be considered for the interpretation of the covariate effects on occurrence of a specific target event of interest. To evaluate the calibration of cumulative incidence functions, \citet{lee2017} generated an alternative kind of calibration plot that compared predictions of the cumulative incidence function to their respective nonparametric estimates.

The simulation study and the analysis of the NP infection data suggest that the methods work well, even in ``unfavorable'' scenarios with a high censoring rate and few type 1 events. However, one should be careful in situations with a large number of time intervals, when the observed number of type 1 events at later time points is rare. 

All evaluations presented in this article were performed using the R add-on package \textbf{discSurv} \citep{discSurv}. It contains the function \texttt{dataLongSubDist()} to generate the binary outcome vectors \eqref{eq:responses} and corresponding weights \eqref{eq:w1} and \eqref{eq:w2}. Parameter estimates of the recalibration model were obtained by using the function \texttt{glm()} with family function \texttt{binomial()} for logistic regression.

\section*{Acknowledgments}
We thank Jan Beyersmann for fruitful discussions on subdistribution hazard modeling and for helpful suggestions on how to improve the manuscript. We thank the SIR-3 study investigators for providing us with the data.
\bibliographystyle{plainnat} 
\bibliography{literatur} 

\newpage

\appendix 

\section{Log-likelihood of the recalibration model \eqref{eq:recal_mod}}\label{app:A}

To derive the log-likelihood of the logistic recalibration model for the discrete subdistribution hazard model, it is assumed that $a=0$, hence the predictor reduces to 
$\eta_{\text{rc}}(t|\bs{x}_i)=b\,\log\left(\hat{\lambda}_1(t|\bs{x}_i)/(1-\hat{\lambda}_1(t|\bs{x}_i))\right)$
and
\begin{align}
\ell_{\text{rc}}\, = &\, \sum_{i=1}^{n}\sum_{t=1}^{k-1}{w_{it}\big\{y_{it} \log(\pi_1(t|\bs{x}_i))+(1-y_{it}) \log(1-\pi_1(t|\bs{x}_i))\big\}} \nonumber \\
=& \, \sum_{i=1}^{n}\sum_{t=1}^{k-1}{w_{it}y_{it}\log\left(\frac{\exp\left(b\,\log\left(\frac{\hat{\lambda}_1(t|\bs{x}_i)}{1-\hat{\lambda}_1(t|\bs{x}_i)}\right)\right)}{1+\exp\left(b\,\log\left(\frac{\hat{\lambda}_1(t|\bs{x}_i)}{1-\hat{\lambda}_1(t|\bs{x}_i)}\right)\right)}\right)} \nonumber \\
&+ \, \sum_{i=1}^{n}\sum_{t=1}^{k-1}{w_{it}(1-y_{it}) \log\left(1-\frac{\exp\left(b\,\log\left(\frac{\hat{\lambda}_1(t|\bs{x}_i)}{1-\hat{\lambda}_1(t|\bs{x}_i)}\right)\right)}{1+\exp\left(b\,\log\left(\frac{\hat{\lambda}_1(t|\bs{x}_i)}{1-\hat{\lambda}_1(t|\bs{x}_i)}\right)\right)}\right)} \nonumber \\
=& \, \sum_{i=1}^{n}\sum_{t=1}^{k-1}{w_{it}y_{it}\log\left(\frac{\frac{\hat{\lambda}_1(t|\bs{x}_i)^b}{(1-\hat{\lambda}_1(t|\bs{x}_i))^b}}{1+\frac{\hat{\lambda}_1(t|\bs{x}_i)^b}{(1-\hat{\lambda}_1(t|\bs{x}_i))^b}}\right)} \nonumber \\
&+ \, \sum_{i=1}^{n}\sum_{t=1}^{k-1}{w_{it}(1-y_{it}) \log\left(\frac{1}{1+\frac{\hat{\lambda}_1(t|\bs{x}_i)^b}{(1-\hat{\lambda}_1(t|\bs{x}_i))^b}}\right)} \nonumber \\
=& \, \sum_{i=1}^{n}\sum_{t=1}^{k-1}{w_{it}y_{it}\log\left(\frac{\hat{\lambda}_1(t|\bs{x}_i)^b}{(1-\hat{\lambda}_1(t|\bs{x}_i))^b}\right)} \nonumber \\
&+ \, \sum_{i=1}^{n}\sum_{t=1}^{k-1}{w_{it}\log\left(1+\frac{\hat{\lambda}_1(t|\bs{x}_i)^b}{(1-\hat{\lambda}_1(t|\bs{x}_i))^b}\right)} \nonumber \\
=& \, b\,\sum_{i=1}^{n}\sum_{t=1}^{k-1}{w_{it}y_{it}\log\left(\hat{\lambda}_1(t|\bs{x}_i)\right)} - \, b\,\sum_{i=1}^{n}\sum_{t=1}^{k-1}{w_{it}y_{it}\log\left(1-\hat{\lambda}_1(t|\bs{x}_i)\right)} \nonumber \\
&- \, \sum_{i=1}^{n}\sum_{t=1}^{k-1}{w_{it}\log\left(\hat{\lambda}_1(t|\bs{x}_i)^b+(1-\hat{\lambda}_1(t|\bs{x}_i))^b\right)} \nonumber \\
&+ \, b\,\sum_{i=1}^{n}\sum_{t=1}^{k-1}{w_{it}\log\left(1-\hat{\lambda}_1(t|\bs{x}_i)\right)} \nonumber \\ 
=& \, b\,\sum_{i=1}^{n}\sum_{t=1}^{k-1}{w_{it}y_{it}\log\left(\hat{\lambda}_1(t|\bs{x}_i)\right)} + \, b\,\sum_{i=1}^{n}\sum_{t=1}^{k-1}{w_{it}(1-y_{it})\log\left(1-\hat{\lambda}_1(t|\bs{x}_i)\right)} \nonumber \\
&- \, \sum_{i=1}^{n}\sum_{t=1}^{k-1}{w_{it}\log\left(\hat{\lambda}_1(t|\bs{x}_i)^b+(1-\hat{\lambda}_1(t|\bs{x}_i))^b\right)}\,. 
\end{align}

\newpage

\section{Further numerical results} \label{app:B}

\begin{figure}[!ht]
\centering
\includegraphics[width=1\textwidth]{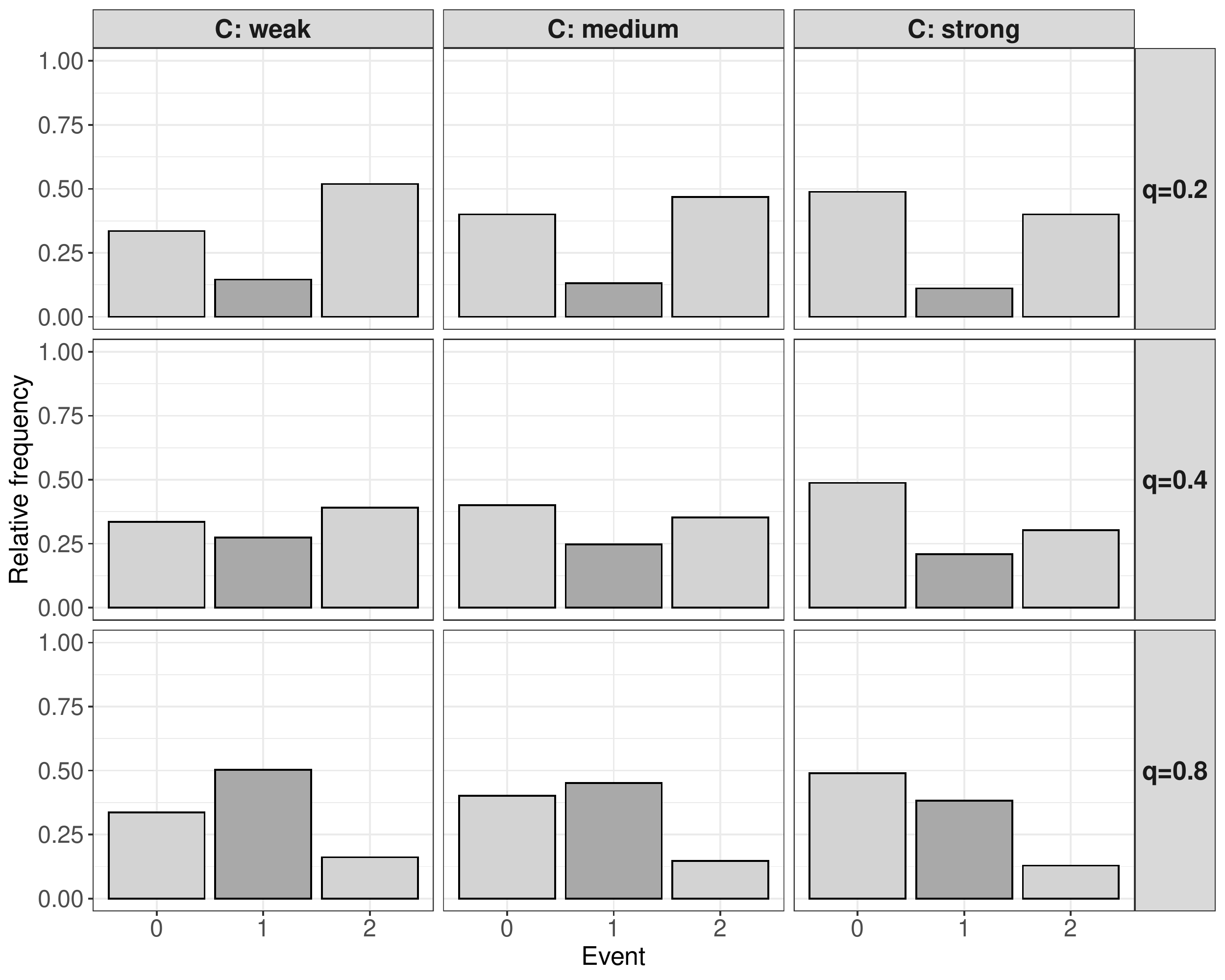}
\caption{Illustration of the experimental design of the simulation study. The bars display the average relative frequencies of observed events (0 = censoring event, 1 = event of interest, 2 = competing event) that were obtained from 100 simulated data sets ($k=5$). The ratio of type~1 and type 2 events was approximately the same in each row  (C = degree of censoring).}
\label{fig:sim_design}
\end{figure}

\begin{figure}[!ht]
\centering
\includegraphics[width=1\textwidth]{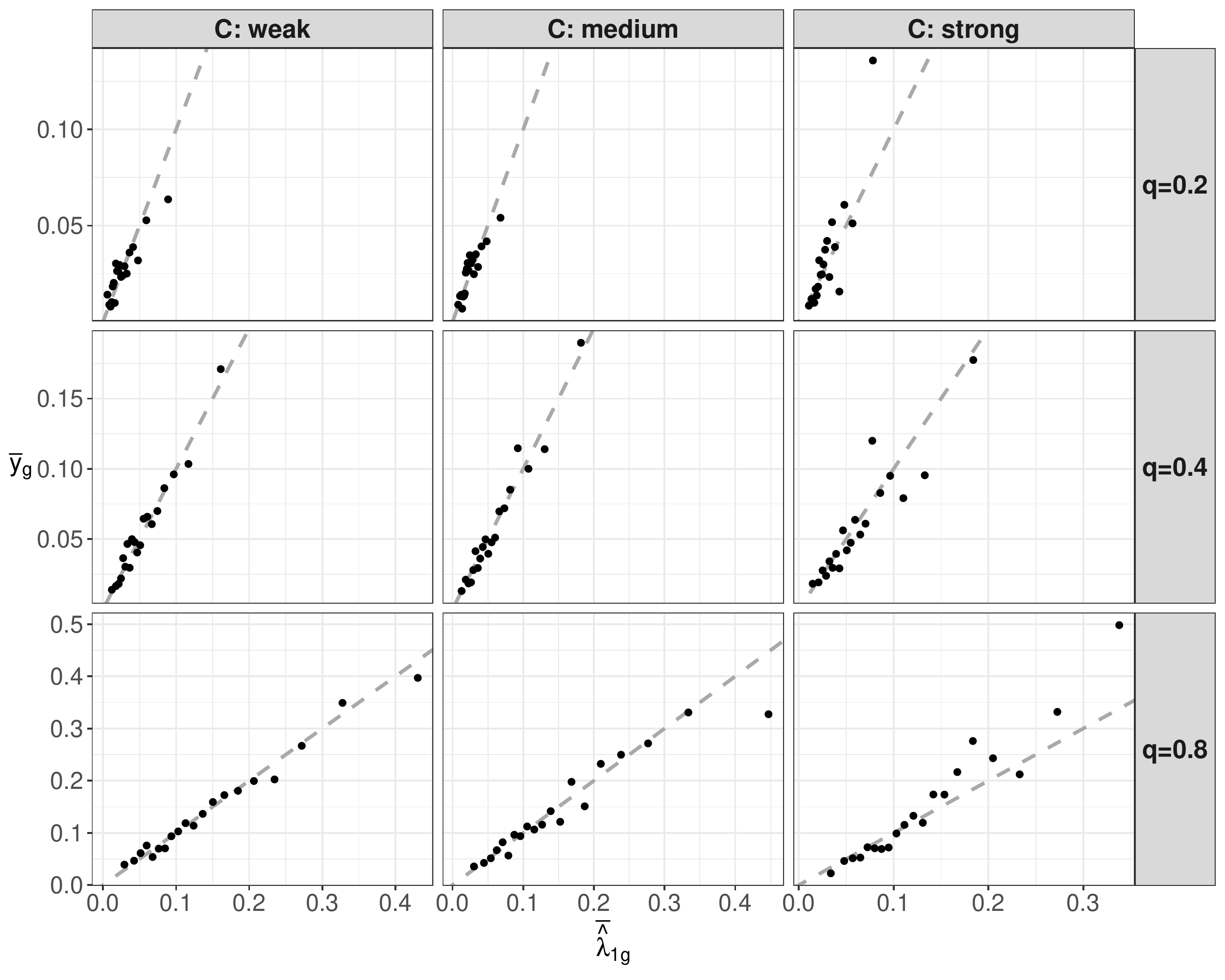}
\caption{Results of the simulation study. Calibration plots of one randomly chosen replication in each simulation scenario using $G=20$ subsets $(k=10)$. The 45-degree lines (dashed) indicate perfect calibration (C = degree of censoring).}
\label{fig:sim_plots10}
\end{figure}

\begin{figure}[!ht]
\centering
\includegraphics[width=1\textwidth]{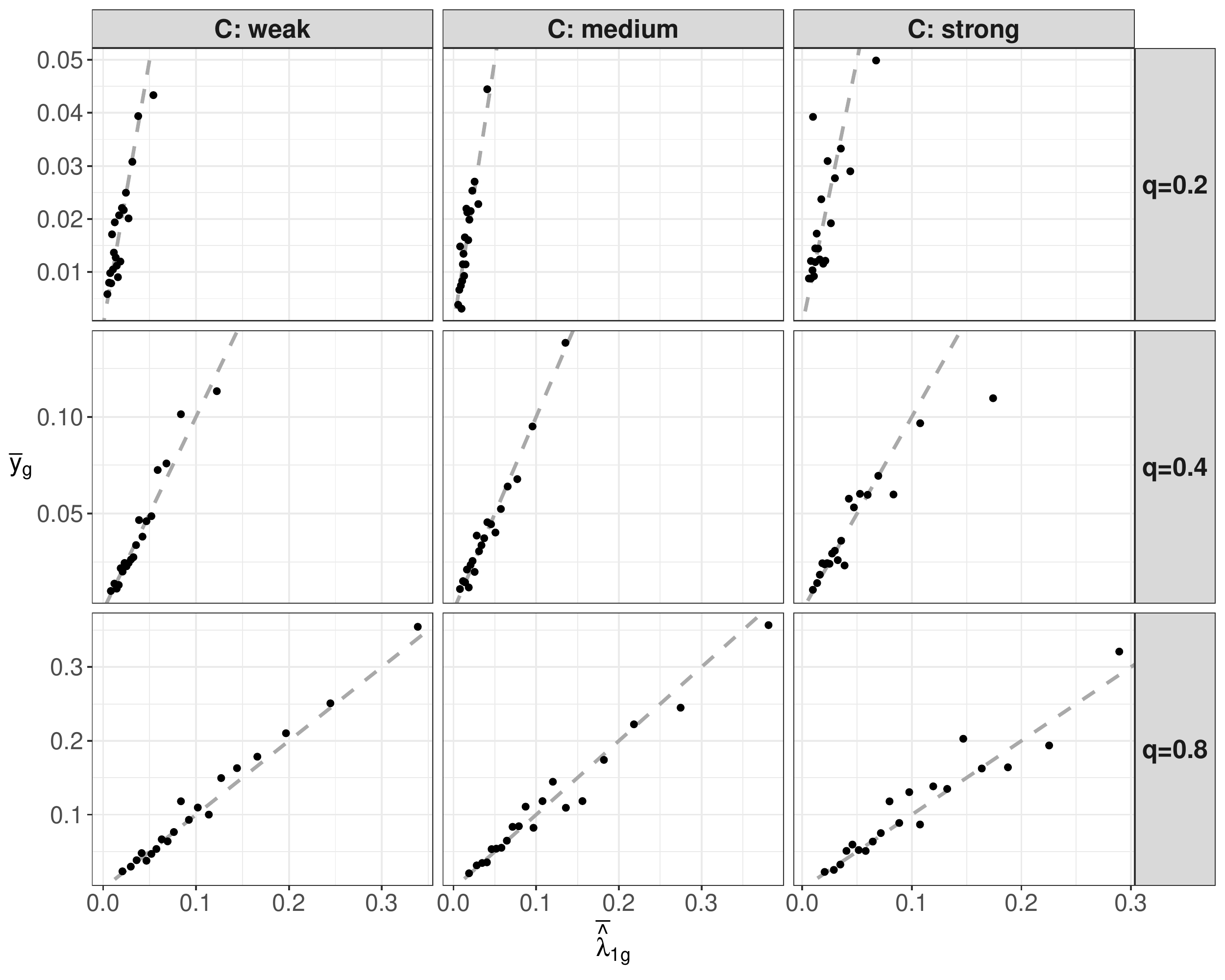}
\caption{Results of the simulation study. Calibration plots of one randomly chosen replication in each simulation scenario using $G=20$ subsets $(k=15)$. The 45-degree lines (dashed) indicate perfect calibration (C = degree of censoring).}
\label{fig:sim_plots15}
\end{figure}

\begin{figure}[!ht]
\centering
\includegraphics[width=1\textwidth]{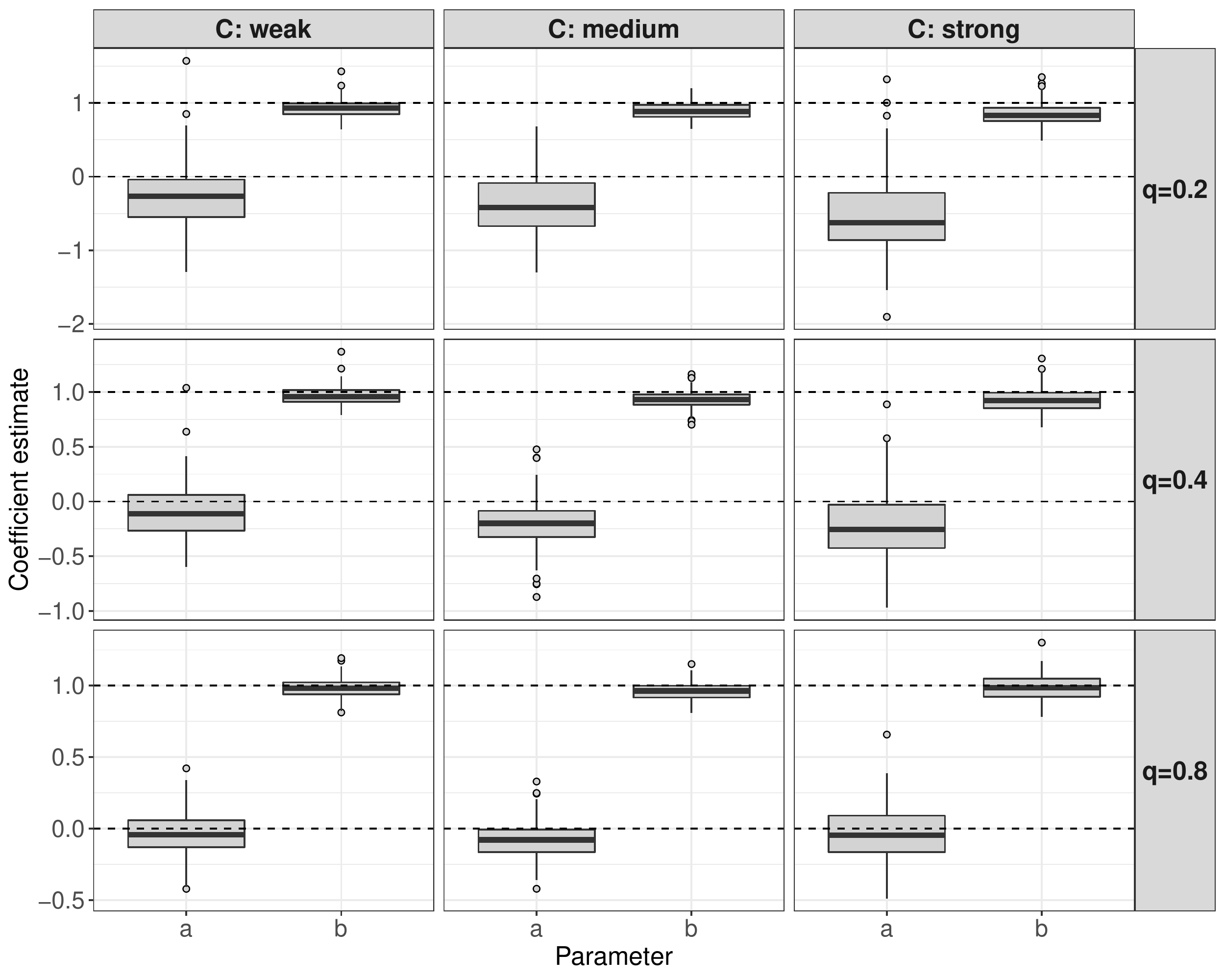}
\caption{Results of the simulation study. The boxplots visualize the estimates of the calibration intercepts $a$ and calibration slopes $b$ that were obtained from fitting the logistic recalibration model $(k=10)$.}
\label{fig:sim_coefs10}
\end{figure}

\begin{figure}[!ht]
\centering
\includegraphics[width=1\textwidth]{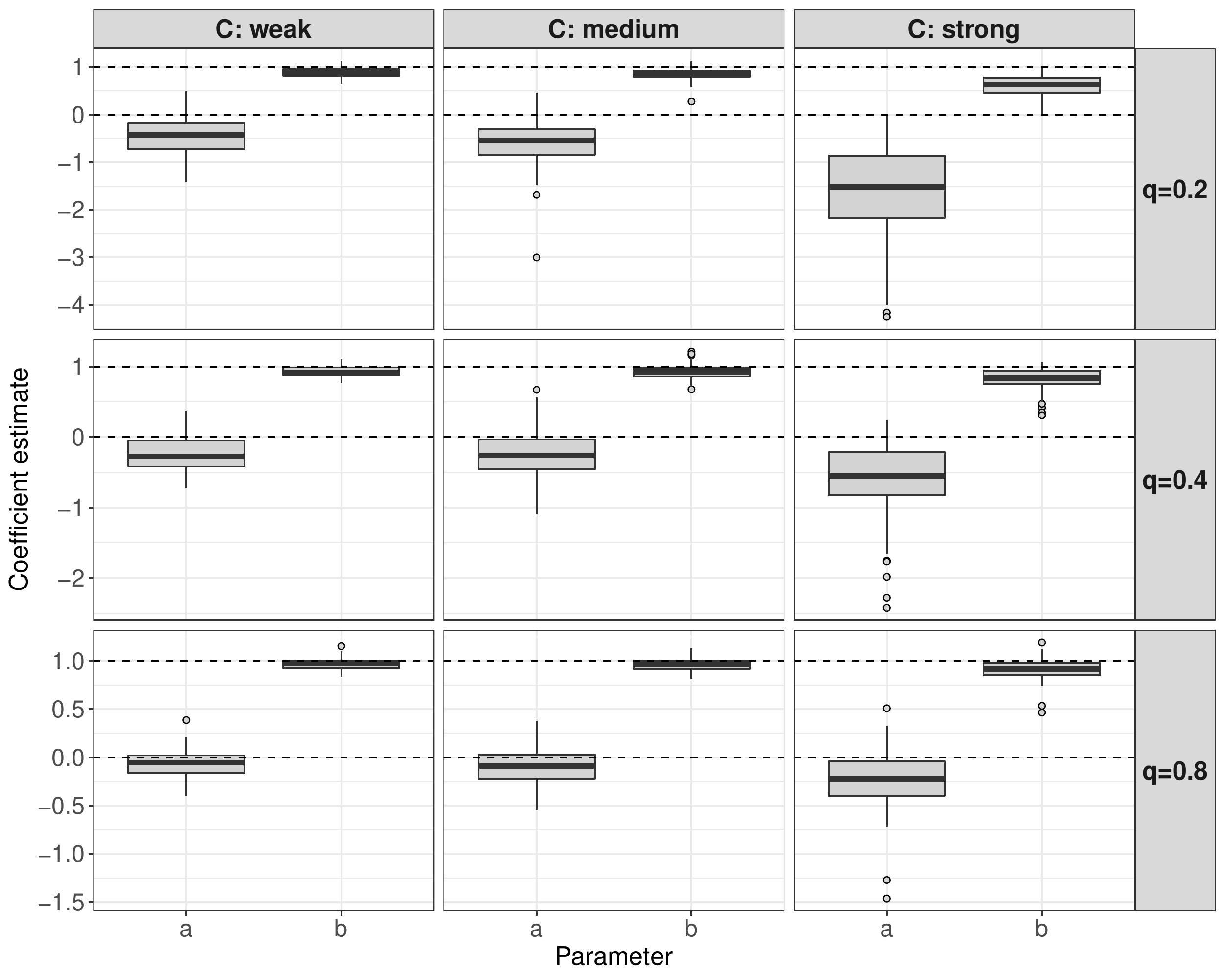}
\caption{Results of the simulation study. The boxplots visualize the estimates of the calibration intercepts $a$ and calibration slopes $b$ that were obtained from fitting the logistic recalibration model $(k=15)$.}
\label{fig:sim_coefs15}
\end{figure}

\begin{figure}[!ht]
\centering
\includegraphics[width=1\textwidth]{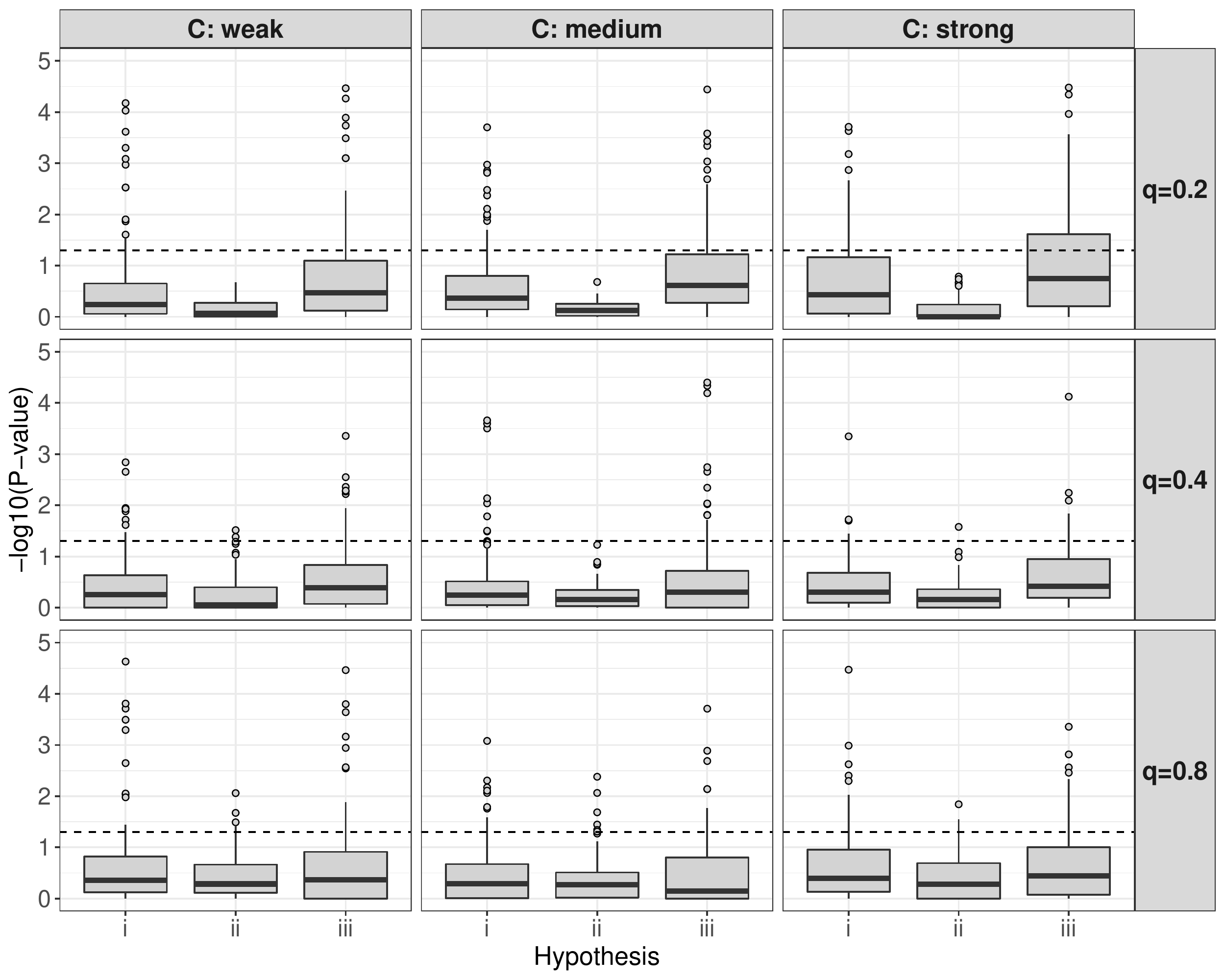}
\caption{Results of the simulation study. The boxplots visualize the negative log10-transformed $p$-values obtained from the recalibration tests~$(k=10)$. The dashed lines correspond to a $p$-value of $0.05$. A value above the dashed line indicates a significant result at the $5\%$ type 1 error level.}
\label{fig:sim_pvalues10}
\end{figure}

\begin{figure}[!ht]
\centering
\includegraphics[width=1\textwidth]{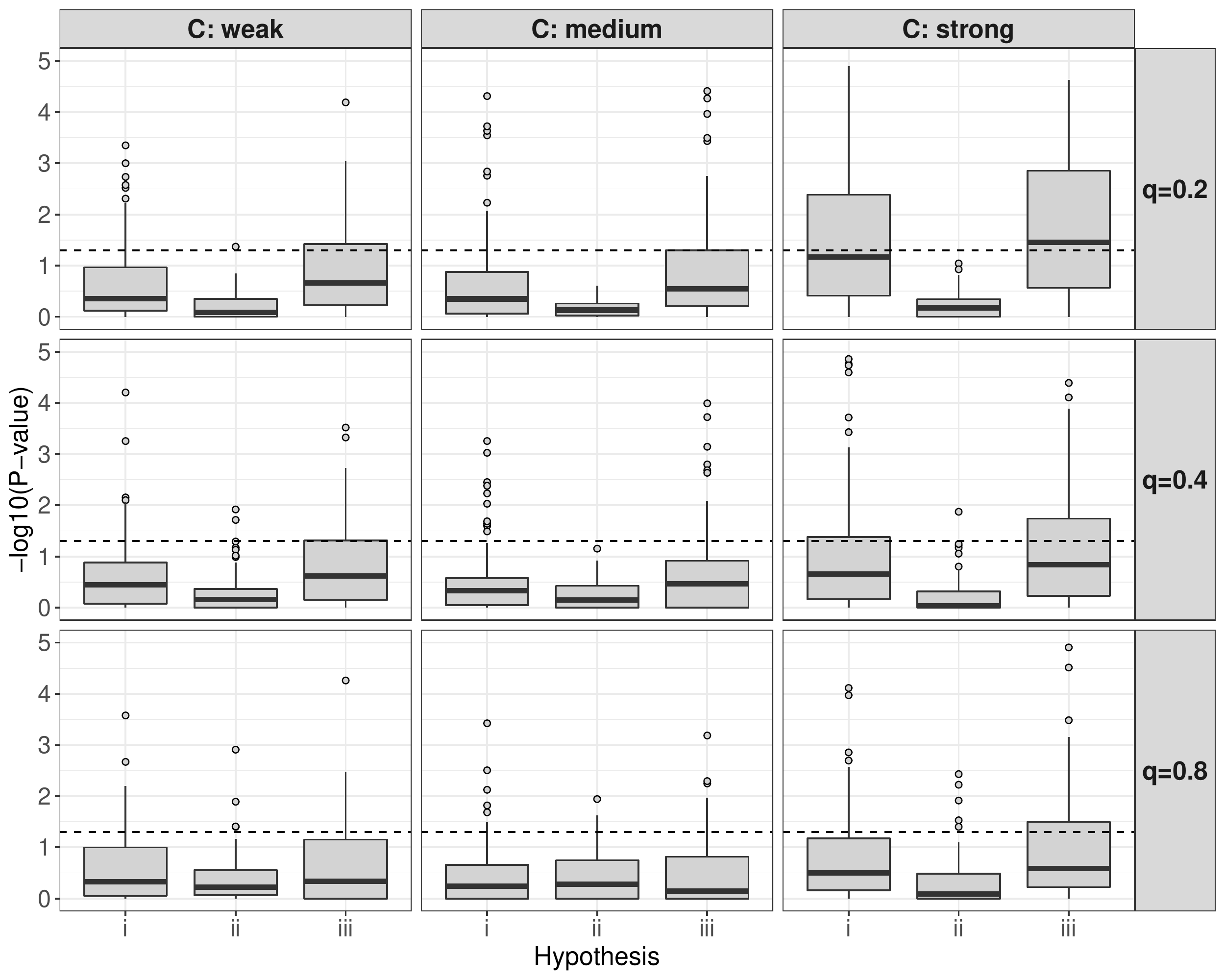}
\caption{Results of the simulation study. The boxplots visualize the negative log10-transformed $p$-values obtained from the recalibration tests~$(k=15)$. The dashed lines correspond to a $p$-value of $0.05$. A value above the dashed line indicates a significant result at the $5\%$ type 1 error level.}
\label{fig:sim_pvalues15}
\end{figure}

\end{document}